\documentclass[11pt]{article}
\usepackage{amsmath,amssymb,color,epsfig,cite}
\usepackage{graphicx}
\usepackage{setspace}
\usepackage{mathrsfs}
\usepackage{commath}
\usepackage[english]{babel}
\usepackage{inputenc}
\usepackage{amsfonts,amsmath,amssymb}
\usepackage{mathrsfs}
\usepackage{nicefrac}
\usepackage{graphicx}
\usepackage{color}
\usepackage{soul} 
\usepackage{hyperref}
\usepackage{accents}
\usepackage{subfigure}
\usepackage[normalem]{ulem}
\usepackage{array}
\usepackage{authblk}
\usepackage{amsfonts}
\newcommand{\hoch}[1]{$\, ^{#1}$}

\makeatletter
\@addtoreset{equation}{section}
\makeatother

\newcommand{\be}{\begin{equation}}
\newcommand{\ee}{\end{equation}}
\newcommand{\bea}{\setlength\arraycolsep{2pt} \begin{eqnarray}}
\newcommand{\eea}{\end{eqnarray}}

\def\0{{\sst{(0)}}}
\def\1{{\sst{(1)}}}
\def\2{{\sst{(2)}}}
\def\3{{\sst{(3)}}}
\def\4{{\sst{(4)}}}
\def\5{{\sst{(5)}}}
\def\6{{\sst{(6)}}}
\def\7{{\sst{(7)}}}
\def\8{{\sst{(8)}}}
\def\sst#1{{\scriptscriptstyle #1}}

\begin{document}

\vspace{15pt}
\begin{center}
{\Large {\bf Generating rotating black hole solutions by using the Cayley-Dickson construction} }

\vspace{15pt}
{\bf Zahra Mirzaiyan\hoch{1} and Giampiero Esposito\hoch{2,1}}

\vspace{10pt}

\hoch{1} {\it INFN Sezione di Napoli,}
\hoch{2}{\it Dipartimento di Fisica ``Ettore Pancini'', \\
Complesso Universitario di Monte S. Angelo, Via Cintia Edificio 6, 80126 Napoli, Italy.}

\vspace{20pt}

\underline{ABSTRACT}

\end{center}
This paper exploits the power of the Cayley-Dickson algebra to generate stationary rotating black hole solutions in one fell swoop. Specifically, we derive the nine-dimensional Myers-Perry solution with four independent angular momenta by using the Janis-Newman algorithm and Giampieri's simplification method, exploiting the octonion algebra. A general formula relating the dimension of the Cayley-Dickson algebra with the maximum number of angular momenta in each dimension is derived. Finally, we discuss the cut-off dimension for using the Cayley-Dickson construction along with the Janis-Newman algorithm for producing the rotating solutions.
\noindent

\thispagestyle{empty}

\vfill
zahra.mirzaiyan@na.infn.it, gesposito@na.infn.it.

\pagebreak
%\voffset=0pt
%\setcounter{page}{1}
%\tableofcontents 
\section{Introduction}
The theory of general relativity describes the dynamical evolution of space-time by presenting a set of non-linear partial differential equations. Solving Einstein's equations to find the exact solutions is a challenging task. Black holes as the famous solutions of the Einstein's equations are important objects theoretically and observationally. They may be the nature's laboratory for quantum gravity. Among the exact solutions of the Einstein's equations, spinning black hole solutions are the most relevant objects to study since it is believed by astrophysicists that black holes in nature are spinning \cite{MP:2011}. Space-time in the exterior regions of rotating stars may be described by these solutions as well. Moreover, there are several reasons for studying rotating solutions of the Einstein's equations in dimensions higher than four \cite{emparan-harvey} like the fact that black holes are related to Lorentzian Ricci-flat manifolds in any dimensions. In addition, string theory obviously requires dimensions higher than four and also contains gravitational systems i.e. black holes \cite{strominger-vafa,JMAPA62023502}. Black holes as gravitational systems with symmetries also have non-trivial connections with soft Weinberg theorem and the gravitational memory effect \cite{stromingerbook}. Since the Weinberg theorem can be investigated in any dimensions, this suggested studying gravitational systems in any dimensions as well. \\\
Janis-Newman approach is known as one of the most mysterious techniques which generates rotating solutions from static ones \cite{JN:1965}. The technique was found in 1965 as an alternative derivation of the Kerr metric. Later, the same approach led to re-derive the charged-rotating black hole solutions in four dimensions known as the Kerr-Newman metric \cite{NCCER:1965}. The approach has been vastly used to generate spinning solutions \cite{tal,Erbin1,Erbin2,Erbin3,Erbin4,Whisker, Schiff,Fer,Kim,emparan-harvey,adam,Azreg1,Azreg2,Azreg3}. By virtue of some difficulties in performing the Janis-Newman algorithm resulting from the use of tetrad bases and the introduction of complex coordinate transformation, a simplification was proposed by Giampieri \cite{Giampieri}. In five dimensions in order to obtain a spinning solution with two independent angular momenta, one has to perform the Janis-Newman algorithm along with double use of Giampieri's simplification to obtain the Myers-Perry solution out of static Schwarzschild one\cite{Erbin1}. Recently, a novel method is proposed for deriving the five dimensional rotating solution using Janis-Newman algorithm and Giampieri's simplification in one fell swoop to avoid the successive use of the Janis-Newman algorithm \cite{mirzaiyan:2017}. The authors exploited the non-commutative property of  the Cayley-Dickson algebra $(n=2)$ \cite{CDalgebra}, i.e. quaternions. Thus, the series of Cayley-Dickson algebras may shed light to obtain higher dimensional rotating solutions with higher number of angular momenta accordingly. The naturally occurring question is whether octonions and sedenions (which are the naturally occurring structures after quaternions) can be exploited to get
further insight, which is the main purpose of the current work. \\
In this paper, taking advantage of the reformulation of the Hopf mapping in terms of rotations, Myers-Perry black holes 
are re-derived in nine dimensions exploiting the octonion algebra and the Janis-Newman approach. It is shown that sedenion algebra also may shed light on the problem in seventeen dimensions. However, due to the cut-off of Cayley-Dickson algebra after sedenions and the non-existence of the Hopf mapping in higher dimensions, the method cannot be used for higher arbitrary dimensions.\\
The paper is organised as follows. In Sec. \ref{Cayley-Dickson}, the Cayley-Dickson algebra is introduced. Myers-Perry black holes together with the Hopf fibration of the corresponding sphere is reviewed in Sec. \ref{MP}. Section \ref{JNsection} is devoted to the Janis-Newman approach. In Sec. \ref{secexamples}, some examples of using the Janis-Newman algorithm are presented. Octonion algebra is introduced in Sec. \ref{9DMPsection} and the Myers-Perry solution in nine dimensions is re-derived. Finally, Sec. \ref{generald} is devoted to the conclusions, and  a general relation between the Cayley-Dickson algebra and the maximum number of angular momenta in each dimension is proposed.

\section{Cayley-Dickson construction}\label{Cayley-Dickson}
The Cayley-Dickson construction produces a series of algebras over the field of real numbers, as a generalisation of complex numbers to higher dimension algebra. The dimension of the algebra is doubled each time one uses the Cayley-Dickson construction \cite{CDalgebra,bocher}. The sequence of the Cayley-Dickson algebras are complex numbers, quaternions, octonions and sedenions which have significant applications in mathematical physics \cite{Carmody,Bilgici,chanyal}. The Cayley-Dickson construction is showing how the Cayley numbers can be constructed as a two-dimensional algebra over quaternions. In fact, starting with a field $F$, the construction yields a sequence of F-algebras of dimension $2^n$ for $n=1,...,4$. We emphasize that there is no algebra after $n=4$. For $n = 2$ it is an associative while non-commutative algebra called a quaternion algebra. For n = 3, it is an alternative while a non-commutative nor associative algebra, called an octonion algebra, and for $n=4$ called as sedenion algebra which lacks the properties of associativity, commutativity and alternativity. Here, we briefly describe how the Cayley-Dickson algebra is constructed.\\
 The Cayley-Dickson algebra is formed by extending a $\ast$-algebra by adding an imaginary unit. We recall that a $\ast$-algebra is an algebra $\mathcal{A}$ upon which a conjugation $a \in \mathcal{A} \rightarrow a^\ast \in \mathcal{A}$ is defined such that
 \begin{eqnarray}\label{staralgebra}
 (a^\ast)^\ast=a, \ \ \ \ (\lambda\ a+\mu\ b)^\ast=\lambda \ a^\ast+\mu \ b^\ast, \ \ \ \ (a b)^\ast=b^\ast\ a^\ast,\ \ \forall a, b \in \mathcal{A},
  \end{eqnarray}
where $\lambda$ and $\mu$ are real. Let $\mathcal{A}$ be a $\ast$-algebra. $\mathcal{A}^\prime$ is defined as
 \begin{eqnarray}
 \mathcal{A} \times \mathcal{A} \equiv \mathcal{A}^\prime= \{a+i b\ |\ a,b \in \mathcal{A}\},
 \end{eqnarray}
 where $i$ is an imaginary unit that does not belong to $\mathcal{A}$. The $\ast$-algebra structure is defined by multiplication and the conjugate of the ordered pair $(a,b)$ as
 \begin{eqnarray}\label{aprime}
&& (a+i b)(c+i d)=(ac-db^{\ast})+i (a^{\ast}d+c b) , \ \ \ \forall a, b, c, d \in \mathcal{A},\nonumber\\
 &&(a+i b)^\ast=a^\ast-i b , \ \ \ \ \ \ \ \ \ \forall a, b \in \mathcal{A}.
 \end{eqnarray}
Such equations define on $\mathcal{A}^\prime$ a $\ast$-algebra structure which extends the one upon $\mathcal{A}$. As a complex number consists of two independent real numbers, they form a two-dimensional vector space over the real numbers which is the case $n=1$ of the Cayley-Dickson construction. The next step of the construction is the passage to quaternions by generalising the multiplication and the conjugate operators. A quaternion can be thought of as a pair of complex numbers and the series may continue to $n=3,$ and $n=4$ as well. \\
 \\The main question we want to address here is that by following the idea of using complex number algebra and quaternions to derive the rotating BTZ black holes and five dimensional Myers-Perry solution from the static stationary Schwarzschild solution, octonions and sedenions can be exploited to get further insight to derive higher dimensional spinning solutions. Using the sequence of Cayley-Dickson algebras and bearing in mind how they help to re-derive the spinning solutions, a general formula is derived which relates the dimension of the algebra, the dimensions of space-time and the maximum number of angular momenta in each dimension.\\\

\section{Spinning black holes in higher dimensions and the Hopf fibration of the n-sphere}\label{MP}
\subsection{Myers-Perry black holes}
The rotating vacuum solutions of the Einstein's equations in dimensions higher than five are known as Myers-Perry solutions.
 For odd dimensions $d=2n+1$ with $d\geq5$ the metric becomes \cite{MP:2011} (with our notation, 
$a_{i}$ and $\mu$ denote the angular momentum and mass parameter of the black hole in
each dimension, while $\mu_{i}$ are the direction cosines)
\begin{eqnarray}\label{MP metric}
&& ds^2=-dt^2+\frac{\mu r^2}{\Pi \ F}(dt+\sum_{i=1}^n \ a_i \ \mu_i^2 \ d\phi_i)^2 +\frac{\Pi \ F}{(\Pi -\mu\ r^2)} dr^2\nonumber\\
&& +\sum_{i=1}^n (r^2+a_i^2)(d\mu_i^2 +\mu_i^2 d\phi_i^2).
\end{eqnarray}
We define 
\begin{eqnarray}\label{pif}
 F=1-\sum_{i=1}^n \frac{a_i^2\ \mu_i^2}{r^2+a_i^2}, \ \ \ \ \ \Pi=\prod_{i=1}^n (r^2+a_i^2).
\end{eqnarray} 
The direction cosines $\mu_i$s are constrained to satisfy
 \begin{eqnarray}\label{mui}
 \sum_{i=1}^n \mu_i^2=1.
 \end{eqnarray}
\\
The main purpose of this paper is to show how one can use the Cayley-Dickson algebra and the Janis-Newman algorithm to derive the rotating solutions defined in Eq. (\ref{MP metric}). 

\subsection{Hopf fibration of the n-sphere}
In this section, we briefly describe the Hopf mapping and its geometric interpretation using rotations \cite{lyson}. To clarify the subject, the example of Hopf fibration of $S^3$ is presented. In this case, the Hopf fibration is the mapping $h:S^3\rightarrow S^2$ defined by
\begin{eqnarray}\label{hopfs3}
h(\alpha,\beta, \gamma,\delta)=(\alpha^2+\beta^2-\gamma^2-\delta^2, 2(\alpha \delta+\beta \gamma), 2 (\beta \delta-\alpha \gamma)),
\end{eqnarray}
such that $(\alpha^2+\beta^2+\gamma^2+\delta^2)=1$. Hamilton's discovery \cite{Hamilton} led to using 4-tuples to make an algebra of rotations in $\mathbb{R}^3$. The invention of quaternions as a set of 4-tuples, was inspired by the generalisation of the rotation of the plane about the origin in two dimensions encoded by complex numbers to an algebra of rotations in $\mathbb{R}^3$ using ordered triplets of real numbers. As a vector space, the set of quaternions is identical to $\mathbb{R}^4$. The vector $(a,b,c,d)$ is written as
\begin{equation}\label{quater2}
 q=a \ e_0+b\ e_1 +c\  e_2+d\  e_3,
\end{equation}
as a quaternion. We shall call the quaternion (\ref{quater2}) a pure quaternion only if $a=0$. The three distinguished coordinate vectors read as
\begin{eqnarray}
e_1=(0,1,0,0), \ \ e_2=(0,0,1,0), \ \ e_3=(0,0,0,1).
\end{eqnarray}
The number $a$ is referred to as the real part and $b,c$, and $d$ are called the $e_1,e_2$ and $e_3$ parts, respectively. The quaternion algebra can be defined as $\mathbb{Q}=(\mathbb{R}^4, +, \Phi)$, which means $\mathbb{R}^4$ supplemented by a set of composition laws. The composition law $\Phi$ for the quaternion algebra $\mathbb{Q}$ is defined by

\begin{eqnarray}
&&\Phi (e_1,e_1)=\Phi (e_2,e_2)=\Phi (e_3,e_3)=-1,\nonumber\\
&&\Phi (e_1,e_2)=e_3, \ \ \Phi (e_2,e_3)=e_1,\ \ \Phi (e_3,e_1)=e_2,\nonumber\\
&&\Phi (e_1,e_2)=-\Phi (e_2,e_1),\ \ \Phi (e_1,e_3)=-\Phi (e_3,e_1),\ \ \Phi (e_2,e_3)=-\Phi (e_3,e_2),\nonumber\\
&&\Phi (e_0,e_1)=\Phi (e_1,e_0),\ \ \Phi (e_0,e_2)=\Phi (e_2,e_0),\ \ \Phi (e_0,e_3)=\Phi (e_3,e_0).
\end{eqnarray}
The conjugate of a quaternion is 
\begin{equation}\label{quatercon}
 q^*=a \ e_0-b\ e_1 -c\  e_2-d\  e_3, 
\end{equation}
which resembles the complex conjugate. Each non-zero quaternion $q$ has a multiplicative inverse, given by 
\begin{eqnarray}
q^{-1}=\frac{q^*}{\abs{q}^2}.
\end{eqnarray}
The quaternions are known as the Cayley-Dickson algebra when $n=2$ and construct an algebra with $D_{\mathbb{Q}}=2^2$ dimensions.\\
Here is how a pure quaternion $q=x\ e_1+y\ e_2+z\ e_3$ determines a linear mapping $Q_q:\mathbb{R}^3\rightarrow \mathbb{R}^3$ as a rotation. Let $P=(x,y,z)$  be a point in 3-space associated to the pure quaternion $q$. The quaternion product $q P q^{-1}$ can be shown to be pure and hence can be thought of as another point $p^\prime=(x^\prime,y^\prime,z^\prime)$ in 3-space. Thus, the linear mapping $Q_q$ is defined by 
\begin{eqnarray}\label{rotation}
Q_q (x, y, z)=q P q^{-1}= (x^\prime,y^\prime,z^\prime).
\end{eqnarray}
Now, let $P_0=(1,0,0)$ be a fixed distinguished point on $S^2$. The Hopf fibration maps the quaterion $q$ to the image of the distinguished point under the rotation
\begin{eqnarray}
q \mapsto Q_q(P_0)= q (1,0,0)q^{-1}.
\end{eqnarray}
It can be shown that the set of points
\begin{eqnarray}\label{circle}
C= (\cos t, \sin t, 0, 0), \ \  t\in \mathbb{R} 
\end{eqnarray}
in $S^3$ all map to the point $(1,0,0)$ on $S^2$ via the Hopf map, which means that each circle as a fiber in $S^3$ is projected to a point in $S^2$. In fact, this set $C$ is the entire set of points that map to the point $(1,0,0)$ via the map $h$. In other words, $C$ is the preimage set $h^{-1} ((1,0,0))$.\\
 A similar construction with the octonions yields the map $h: S^7\rightarrow S^4$, and $h:S^{15}\rightarrow S^8$ using sedenions, which shows that a reformulation of Hopf fibration in terms of rotations is possible in $S^7$ and $S^{15}$. It should be noted that the Hopf fibration can only occur for the mentioned spheres according to the Adam's theorem \cite{adam}.\\
The Hopf fibration of the n-sphere and its connection to the Cayley-Dickson field algebras is discussed in \cite{Hatsuda:2009}, where a simple derivation of metrics for Hopf fibrations by a coset formulation is presented. Specifically, the Hopf fibration for a round $S^{4N+3}$ leads to a metric which reads as
\begin{eqnarray}\label{nsph}
ds^2_{S^{4N+3}}=d\theta^2+\sin^2\theta\ d\Omega^2_{(4N-1)}+\cos^2\theta\ d\Omega^2_{3},
\end{eqnarray}
\\
where the metric $d\Omega^2_3$ is given by

\begin{eqnarray}\label{3sph}
d\Omega^2_3=d\Theta^2+ \sin^2\Theta\ d\Phi_1^2+\cos^2\Theta\ d\Phi_2^2.
\end{eqnarray}
\\
The Hopf fibration of the n-sphere, Eqs. (\ref{nsph}) and (\ref{3sph}) will be used to define the metric on the 3-sphere and 7-sphere in the proceeding sections. It should be emphasised that the Hopf fibration of the n-sphere in Janis-Newman approach using Cayley-Dickson algebra is an essential step.

\section{Janis-Newman algorithm and Giampieri's simplification}\label{JNsection}
In this section, we review the Janis-Newman algorithm which is used to derive the rotating  black hole solutions from the static Schwarzschild metric. The Janis-Newman approach is based on introducing a set of null tetrads and a series of complex conjugation transformations  proposed in 1965 \cite{NJ}. One can easily formulate the algorithm for deriving the rotating solutions with the following steps:\\\

\noindent 1. Transforming the metric which is written in the Boyer-Lindquist (BL) coordinates to the Eddington-Finkelstein (EF) coordinates by using the following transformation:
\begin{eqnarray}\label{BLtoED}
u=t-r_{*}, \ \ \ \ r_{*}=\int\ dr \sqrt{\frac{-g_{rr}}{g_{tt}}}.
\end{eqnarray}

\noindent 2. Writing the contravariant metric in terms of a set of null tetrads $l_\mu, n_\mu,m_\mu,\bar{m}_\mu$ such that $l_\mu n^\mu=-m_\mu \bar{m}^\mu=-1$ ($l^\mu$ and $n^\nu$ are considered to be real-valued components while $m^\mu$ is the complex conjugate of $\bar{m}^\mu$) as
\begin{eqnarray}\label{metricin tetrad}
g^{\mu\nu}=l^\mu n^\nu+n^\mu l^\nu-m^\mu\bar{m}^\nu-\bar{m}^\mu m^\nu.
\end{eqnarray}

\noindent 3. Complex conjugate transformations of the coordinates $u$ and $r$ while introducing new parameters which are related to the angular momentum parameters in the final solution.\\

\noindent 4. Rewriting the new sets of tetrads and obtaining the rotating metric in the Eddington-Finkelstein coordinates.\\

\noindent 5. Reverting to the Boyer-Lindquist coordinates.\\
\\
One can avoid null tetrads by using the simplification proposed by G. Giampieri in 1990 \cite{Giampieri}. Giampieri's method is based on applying some angle-fixing ansatz instead of introducing the sets of null tetrads. For some explicit examples in various dimensions see \cite{Erbin1,Erbin2,Erbin3,Erbin4,mirzaiyan:2017}.  Some of the examples are reviewed in section \ref{secexamples}.

%@@@@@@@@@@@@@@@@@@@@@@@@@@@@@@@@@@@@@@@@@

\section{Examples of Cayley-Dickson algebra and Janis-Newman approach for building rotating solutions}\label{secexamples}
In this section, we review the derivation of spinning BTZ solution in three dimensions by using the Janis-Newman algorithm and exploiting the Cayley-Dickson algebra with $d=2^n$ with $n=1$ algebra\cite{Erbin3}. Then we switch to the case of spinning five-dimensional Myers-Perry while the quaternion algebra is used, which is the Cayley-Dickson algebra with $n=2$, discussed in \cite{mirzaiyan:2017}.

%@@@@@@@@@@@@@@@

\subsection{Using complex numbers for generating rotating BTZ black holes in three dimensions}\label{BTZsection}
We consider the non-rotating BTZ black hole in coordinates $(t,r, \phi)$ as 

\begin{eqnarray}\label{BTZBL}
ds^2=-f(r)dt^2+\frac{dr^2}{f(r)}+r^2\ d\phi^2, \ \ \ \ f(r)=-m+\frac{r^2}{l^2},
\end{eqnarray}
with $m$ and $l$ defined as the mass and the radius of $AdS_3$ space. It should be noted that in this case the angular part of the metric (\ref{BTZBL}) is a 1-sphere ($S^1$). One can write the metric (\ref{BTZBL}) in the Eddington-Finkelstein coordinates as

\begin{eqnarray}\label{BTZEF}
ds^2=-du^2-2du\ dr+r^2 d\phi^2.
\end{eqnarray}
Hence, introducing the new parameter $\mu$ with the constraint $\mu^2=1$, the null coordinates $u$ and $r$ can be complexified and transformed as 
\begin{eqnarray}\label{BTZur}
u=u^\prime+i\ a \sqrt{1-\mu^2},\nonumber\\
r=r^\prime-i\ a \sqrt{1-\mu^2}.\nonumber\\
\end{eqnarray}
Thus, the metric (\ref{BTZEF}) transforms as
\begin{eqnarray}
&&ds^2=-du^2-2du\ dr+(r^2+a^2)(d\mu^2+\mu^2\ d\phi^2)-2a\ \mu^2\ dr d\phi\nonumber\\
&&+(1-\tilde{f})(du+a\mu^2d\phi^2).
\end{eqnarray}
\\
The transformation of $f$ reads as 
\begin{eqnarray}
\tilde{f}=-m+\frac{r^2+a^2(1-\mu^2)}{l^2}\equiv -m+\frac{\rho^2}{l^2} .
\end{eqnarray}
\\
Upon using the following coordinate transformations the metric can be re-expressed in the Boyer-Lindquist coordinates:

\begin{eqnarray}
&&du=dr-g(r)dr,\ \ \ d\phi=d\phi^\prime -h(r) dr,\nonumber\\
&&g(r)=\rho^2 (1-\tilde{f}), \ h(r)=\frac{a}{\Delta}, \ \Delta=r^2+a^2+(\tilde{f}-1)\rho^2.
\end{eqnarray}
Fixing $\mu^2=1$, the spinning BTZ solution can be derived as (primes are deleted)
\begin{eqnarray}
&&ds^2=-N^2 dt^2+N^{-2} dr^2+r^2 (N^{\phi}dt+d\phi)^2,\nonumber\\
&&N^2=-m+\frac{r^2}{l^2}+\frac{a^2}{r^2}, \ \ N^{\phi}=\frac{a}{r^2}.
\end{eqnarray}
The spinning charged BTZ solution can be also derived by using the same method, starting from a charged static BTZ black hole metric \cite{Erbin2}.

%@@@@@@@@@@@@@@@@@@@@@@@@@@@@

\subsection{Using quaternions for producing rotating Myers-Perry black hole in five dimensions}\label{5DMPsection}
In this section, we review how the five-dimensional Myers-Perry black hole can be derived exploiting quaternions algebra and Janis-Newman algorithm along with Giampieri's simplification in one fell swoop instead of successive use of the Janis-Newman formalism \cite{mirzaiyan:2017}.
\\
The Schwarzschild metric in five dimensions in $(t,r,\theta,\phi_1,\phi_2)$ coordinates can be written as 

 \begin{equation}\label{5sch}
 ds^{2}=-f(r) dt^{2}+{f(r)}^{-1} dr^2+r^{2} d{\Omega}^{2} _{3}, \ \ \ \ f(r)=1-\frac{m}{r^2},
 \end{equation}
 where $d\Omega_3^2$ is the metric induced on the sphere $S^{3}$  in Hopf coordinates described in Eq. (\ref{3sph}).
 The metric (\ref{5sch}) in the Eddington-Finkelstein coordinates can be written as
  \begin{equation}\label{newm}
 ds^{2}=-du (du+2dr)+(1-f(r)) du^{2}+r^{2} d{\Omega}^{2} _{3}.
 \end{equation}
 On introducing $\chi_1$ and $\chi_2$ as two angles to help us performing the simplification method, one can propose the following transformations for the null coordinates $u$ and $r$ to perform the complex transformation step as:
 \begin{eqnarray}\label{newtr}
 u&=& u^{\prime}+e_1 a_1\ \cos{{\chi}_{1}}+e_2 a_2\ \sin {{\chi}_{2}},\nonumber\\
 r&=& r^{\prime}-e_1 a_1\ \cos{{\chi}_{1}}-e_2 a_2\ \sin {{\chi}_{2}}.\nonumber\\
  \end{eqnarray}
  With our notation $a_1$ and $a_2$ are two parameters which are related to independent angular momenta. The transformation takes place in both $(r,\phi_1)$ and $(r,\phi_2)$ simultaneously. Two angles $\chi_1$ and $\chi_2$ will be fixed by the angle-fixing ansatz
  \begin{eqnarray}\label{GA}
  e_1 d{{\chi}_{1}}&=&\sin{{\chi}_{1}} \ d{\phi_1},\ \ {{\chi}_{1}}={\theta},\nonumber\\
  e_2 d{{\chi}_{2}}&=&-\cos{{\chi}_{2}} \ d{\phi_2},\ \ {{\chi}_{2}}={\theta}.
\end{eqnarray}
\\
With the set of transformations (\ref{newtr}), taking advantage of the associativity property of quaternions, we have 
\begin{eqnarray}\label{rr}
r^{2}=r r^{*}=r^{{\prime}{2}} +a_1^{2} \cos^{2} {\theta}+a_2^{2} \sin^{2} {\theta}.
\end{eqnarray}
Therefore, the functional form of $f(r)$ under the complexification transformations can be derived as 
\begin{equation}\label{newf}
f(r^{\prime})=1-\frac{m}{r^{{\prime}{2}} +a_1^{2} \cos^{2} {\theta}+a_2^{2} \sin^{2} {\theta}}.
\end{equation}
Thus, under the transformations (\ref{newtr}), the metric (\ref{newm}) takes the form

\begin{eqnarray}\label{trametric} 
ds^{2}&=&-(du^{\prime}-e_1 a_1\ \sin{{\chi}_{1}}\ d{{\chi}_{1}} +e_2 a_2\ \cos{{\chi}_{2}}\ d{{\chi}_{2}}) \times \nonumber\\
&& [((du^{\prime}-e_1 a_1\ \sin{{\chi}_{1}}\ d{{\chi}_{1}} +e_2 a_2 \ \cos{{\chi}_{2}}\ d{{\chi}_{2}})+2(dr^{\prime}\nonumber\\
&&+e_1 a_1\ \sin{{\chi}_{1}}\ d{{\chi}_{1}} - e_2 a_2\ \cos{{\chi}_{2}}\ d{{\chi}_{2}}))]\nonumber\\
&& +(1-{\tilde{f}}(r^{\prime})) (du^{\prime}-e_1 a_1\ \sin{{\chi}_{1}} d{{\chi}_{1}} +e_2 a_2 \ \cos{{\chi}_{2}}\ d{{\chi}_{2}})^{2}\nonumber\\
&& +\mbox{angular part of the metric}.
\end{eqnarray}
In the angular momentum part, the following terms have to be transformed.\\
\\
\noindent 1.The term $r^{2} d{\theta}^{2}$
\begin{equation}\label{ft}
r^{2} d{\theta}^{2}\longrightarrow (r d\theta) ({r d\theta)^*} = (r^{{\prime}{2}} +a_1^{2} \cos^{2} {\theta}+a_2^{2} \sin^{2} {\theta})d{\theta}^{2}.
\end{equation}

\noindent 2. The second term is $r^{2} \sin^{2} {\theta} d{\phi_1}^{2}$.
Here, we have the angle $\phi_1$ and   we know from the angle-fixing conditions in (\ref{GA}) that  the  transformation in the $(r,\phi_1)$ plane should be accomplished via angle ${\chi}_{1}$
 \begin{eqnarray}\label{st}
 r^{2} \sin^{2} {\theta} d{\phi_1}^{2}\longrightarrow \sin^{2} {\theta} (r d{\phi_1}) (r d{\phi_1})^{\ast}.
 \end{eqnarray}
 We now proceed with the calculation of  the term $r d {\phi}_1$ by replacing it with the following ansatz (see Eq. (\ref{GA})):
 \begin{eqnarray}\label{dphi}
 &&r d{\phi_1}= r \ \frac{e_1 d{{\chi}_{1}}}{\sin{{\chi}_{1}}}=\Bigl(\frac{e_1 \cdot r+r\cdot e_1}{2}\Bigl)\ \frac{ d{{\chi}_{1}}}{\sin{{\chi}_{1}}}= \nonumber\\
 &&\biggr[ \frac{e_1\cdot (r^{\prime}-e_1 a_1\  \cos{{\chi}_{1}}-e_2 a_2\ \sin {{\chi}_{2}})+(r^{\prime}-e_1 a_1\  \cos{{\chi}_{1}}-e_2 a_2\ \sin {{\chi}_{2}})\cdot e_1}{2} \nonumber\\
 &&\times \frac{d{{\chi}_{1}}}{\sin{{\chi}_{1}}}\biggr] \nonumber\\
 &&=\frac{e_1 d{{\chi}_{1}}}{\sin{{\chi}_{1}}} ( r^{\prime} -e_1 a_1\ \cos{{\chi}_{1}}).
 \end{eqnarray}
 It should be noted that $e_1$ and $e_2$ are non-commutative and therefore  we use a symmetric form in all products to have meaningful relations. The  last line of the above relation was obtained by using the non-commutative feature of quaternion multiplication $(e_1 \cdot e_2=-e_2 \cdot e_1)$.\\
  The transformed   form of (\ref{st}) now reads as follows:
  \begin{eqnarray}\label{st2}
  && r^{2} \sin^{2} {\theta} d{\phi_1}^{2}\longrightarrow \nonumber\\
 &&  \ \  \sin^{2} {\theta}\ \ ( r^{\prime} -i a_1\ \cos{{\chi}_{1}}) ( r^{\prime} + ia_1\ \cos{{\chi}_{1}})\  \frac{ d{{\chi}_{1}^{2} }}{\sin^{2} {{\chi}_{1}}}\nonumber\\
 &&= \sin^{2} {\theta}  (r^{{\prime}{2}} + a_1^2 \cos^{2} {\theta})\ d{{\phi_1}^{2}},
  \end{eqnarray}
where  $d{\phi_1}^2=\frac{ d{{\chi}_{1}^{2} }}{\sin^{2} {{\chi}_{1}}}$ is used on the third line and ${\chi}_{1}=\theta$ in the last line which is the angle-fixing condition in (\ref{GA}). The basic point in this method is the replacement of the products of non-commutative quaternions with a symmetric form.\\

\noindent 3. The third  term is $r^{2} \cos^{2} {\theta}\  d{\phi_2}^{2}$. One can use a discussion similar to the  transformation method of the second term. Thus the third term in the angular part of the metric (\ref{newm}) transforms as 
  \begin{eqnarray}\label{tt2}
 r^{2} \cos^{2} {\theta}\ d{\phi_2}^{2}\longrightarrow \cos^{2} {\theta}\ (r^{{\prime}{2}} +a_2^{2} \sin^{2} {\theta})\ d{{\phi_2}^{2}}.
 \end{eqnarray}
 \\
 Therefore the transformed metric in the Eddington-Finklestein coordinates is derived as
 \begin{eqnarray}\label{lastmetric}
ds^{2}&=&-du^{2} -2 du dr \nonumber\\
&&+(1-\tilde{f} {(r)}) (du-a_1\ \sin^{2} {\theta}\ d{\phi} -a_2\ \cos^{2}{\theta}\ d{\psi} )^{2} \nonumber\\
&&+2a_1\ \sin^{2} {\theta}\ dr d{\phi}+2a_2\ \cos^{2} {\theta}\ dr d{\psi}+ {\rho}^{2} d{\theta}^{2}\nonumber\\
&&+(r^{2} +a_1^{2}) \sin^{2} {\theta}\ d{\phi_1}^{2}+(r^{2} +a_2^{2}) \cos^{2} {\theta}\ d{\phi_2}^{2}, \
\end{eqnarray}
where
\begin{equation}\label{ro}
{\rho}^{2}=r^{2} +a_1^{2} \cos^{2} {\theta} +a_2^{2} \sin^{2} {\theta}.
\end{equation}
\\
The metric can be re-expressed in the Boyer-Lindquist coordinates with the following transformations:
\begin{eqnarray}\label{BLT}
du&=& dt-g(r) dr,\nonumber\\
d\phi_i &=& d{{\phi_i}^{\prime}}- h_{\phi_i}(r) dr,
\end{eqnarray}
where
\begin{eqnarray}
g(r)&=&\frac{\Pi}{\Delta},\nonumber\\
h_{\phi_i}(r)&=&\frac{\Pi}{\Delta} \frac{a_i}{r^{2}+a_i^{2}},
\end{eqnarray}
\\
based on the definition
\begin{eqnarray}
\Pi &=&({r^{2}+a_1^{2}})({r^{2}+a_2^{2}}),\nonumber\\
\Delta &=&r^{4} + r^{2} (a_1^{2} +a_2^{2}-m)+a_1^{2} a_2^{2}.
\end{eqnarray}
 Finally, we can find the Myers-Perry metric in five dimensions in the form
 \begin{eqnarray}\label{fmetric}
 ds^{2}&=&-dt^{2} +(1-\tilde{f}(r))(dt-a_1\ \sin^{2} {\theta}\ d{\phi_1} -a_2\ \cos^{2}{\theta}\ d{\phi_2} )^{2}
\nonumber\\
 &&+{\rho}^2 d{\theta}^{2} +\frac{r^{2} {\rho}^{2}}{\Delta} dr^2+(r^{2} +a_1^{2}) \sin^{2} {\theta}\ d{\phi_1}^{2} \nonumber\\
&&+(r^{2}+a_2^{2}) \cos^{2} {\theta}\ d{\phi_2}^{2}.
 \end{eqnarray}
Therefore, the Myers-Perry solution in five dimensions is derived by using the non-commutative property of quaternion algebra and the Janis-Newman approach in one fell swoop. It should be emphasised that the power of Cayley-Dickson algebra with $n=2$ prevents us from successive use of Janis-Newman approach.

\section{Exploiting higher dimensional algebra}\label{9DMPsection}
\subsection{Octonions}

The octonions can be thought of as octets (or 8-tuples) of real numbers. Octonions shed light on the rotation of the plane about the origin in $\mathbb{R}^7$ space. Every octonion is a real linear combination of the unit octonions:\\
\begin{eqnarray}\label{octbasis}
\{e_0, e_1, e_2, e_3, e_4, e_5, e_6, e_7\},
\end{eqnarray}
where $e_0$ is the scalar or real element and it may be simply identified with a real number. Octonions are generally represented in a specific form as follows:

\begin{equation}\label{quater}
 o= a\ e_0+b\ e_1 +c\  e_2+d\  e_3+ f\ e_4+ g\ e_5+ h\ e_6+ l\ e_7,
\end{equation}
with its conjugate $o^*$ defined as
\begin{equation}\label{quater}
o^*= a\ e_0-b\ e_1 -c\  e_2-d\  e_3- f\ e_4- g\ e_5- h\ e_6- l\ e_7.
\end{equation}
The product of an octonion with its conjugate, $o^*o = oo^*$, is always a non-negative real number:
\begin{eqnarray}
o^*o=a^2+b^2+c^2+d^2+f^2+g^2+h^2+l^2,
\end{eqnarray}
where, $a, b, c, d, f, g, h,$ and $l$ are real numbers and $e_0, e_1, e_2, e_3, e_4, e_5, e_6,$ 
and $e_7$ are octonion units defined as the following coordinate vectors:
\begin{eqnarray}
&&e_1=(0,1,0,0,0,0,0,0), \ \ e_2=(0,0,1,0,0,0,0,0),\ \ e_3=(0,0,0,1,0,0,0,0),\nonumber\\
&&e_4=(0,0,0,0,1,0,0,0),\ \ e_5=(0,0,0,0,0,1,0,0), \ \  e_6=(0,0,0,0,0,0,1,0),\nonumber\\
&&e_7=(0,0,0,0,0,0,0,1).
\end{eqnarray}
Whereas the multiplication rules for the octonion algebra $\mathbb{O}$ are defined as (with $i, j, k=1,...,7)$
\begin{eqnarray}
&&e_0=1,\\
&&\Phi(e_0,e_i)=\Phi(e_i,e_0),\nonumber\\
&&\Phi(e_i,e_i)=-1,\nonumber\\
&&\Phi(e_i,e_j)=-\Phi(e_j,e_i)=-\delta_{ij}+\gamma_{ijk} e_k.\nonumber\\
\end{eqnarray}
The structure constants $\gamma_{ijk}$ are antisymmetric and define the following triplets:
\begin{eqnarray}
&&\gamma_{ijk}=+1,\ \  \  \ \forall (i,j,k) \ \ \text{equal to}\nonumber\\
&&(1,2,3), (1,4,5),(1,7,6),(2,3,1), (2,4,6), (2,5,7),(3,1,2), (3,4,7), \nonumber\\
&&(3,6,7), (4,5,1), (4,6,2), (4,7,3), (5,1,4), (5,3,6), (5,7,2), (6,1,7), \nonumber\\
&& (6,2,4), (6,5,3), (7,2,5), (7,3,4), (7,6,1).
\end{eqnarray} 
\\
Using $e_0, e_1, e_2, e_3, e_4, e_5, e_6$, and $e_7$ as the base which results in a 8-axis, the octonions construct a space with $D_{\mathbb{O}}=2^3$ dimensions as a Cayley-Dickson algebra with $n=3$. We emphasise that the octonion algebra lacks the features of associativity and commutativity. Similar to the discussion of the Hopf map of $S^3$, the Hopf fibration of $S^7$ can be reformulated as rotations in $S^7$ in terms of octonions.

%@@@@@@@@@@@@@@@@@@@@@@@@@@@@
\subsection{Generating nine-dimensional rotating solutions using octonions}
In this section, we exploit the octonion algebra to derive the Myers-Perry solution in nine dimensions in one fell swoop. It should be noted that one can take advantage of  the successive use of Janis-Newman algorithm. However, in this approach the complex number system is sufficient. Octonion algebra makes the process easier and the transformation can be performed in one step. The nine-dimensional Schwarzschild metric in the Edington-Finklestein retarded null coordinates reads as 
\begin{eqnarray}\label{null9dm}
ds^2=-du\ (du+2dr)+(1-f(r)) \ du^2 +r^{2} d{\Omega}^{2} _{7},\ \ \ F=1-\frac{2m}{r^6},
\end{eqnarray}
where $d\Omega^2_7$ is the induced metric on the 7-sphere. 
The Hopf fibration of 7-sphere reads as 
\begin{eqnarray}\label{hopf7}
&&ds^2_7=d\theta_1^2+\sin^2\theta_1\ d\theta_2^2+\sin^2\theta_1\ \sin^2\theta_2\ d\phi_1^2+\sin^2\theta_1\ \cos^2\theta_2\ d\phi_2^2\nonumber\\
&&+\cos^2\theta_1\ d\theta_3^2+\cos^2\theta_1\ \sin^2\theta_3\ d\phi_3^2+\cos^2\theta_1\ \cos^2\theta_3\ d\phi_4^2.
\end{eqnarray}
Thus, the direction cosines $\mu_i$s $(i=1,..,4)$ are defined as (angles on the 7-sphere are $\theta_1,\theta_2, \theta_3 \in [0,\pi/2] $ and $\phi_1, \phi_2,\phi_3, \phi_4 \in [0,2\pi] $.)
  \begin{eqnarray}\label{mu9}
 \mu_1=\sin\theta_1\ \sin\theta_2,\ \ \ \mu_2=\sin\theta_1\ \cos\theta_2,\ \ \ \mu_3=\cos\theta_1\ \sin\theta_3,\ \ \ \mu_4=\cos\theta_1\ \cos\theta_3.
 \end{eqnarray}
 
 \subsection{Transformations of the null coordinates $u$ and $r$}
 
The transformation laws for the $u$ and $r$ coordinates are proposed as

\begin{eqnarray}\label{rutras}
&&r=r^\prime -e_1\ a_1\ \cos\chi_1\ \sin\chi_2-e_2\ a_2\ \sin\chi_1\ \sin\chi_2-e_3\ a_3\ \sin\chi_3 \ \sin\chi_4\nonumber\\
&&-e_4\ a_4\ \cos\chi_3\ \sin\chi_4,\nonumber\\
&&\equiv r^\prime -e_1\ a_1\ \mathcal{F}(\chi_1,\chi_2)-e_2\ a_2\ \mathcal{G}(\chi_1,\chi_2)-e_3\ a_3\ \mathcal{H}(\chi_3,\chi_4)-e_4\ a_4\ \mathcal{M}(\chi_3,\chi_4),\nonumber\\
&&u=u^\prime +e_1\ a_1\ \cos\chi_1\ \sin\chi_2+e_2\ a_2\ \sin\chi_1\ \sin\chi_2+e_3\ a_3\ \sin\chi_3 \ \sin\chi_4\nonumber\\
&&+e_4\ a_4\ \cos\chi_3\ \sin\chi_4,\nonumber\\
&&\equiv u^\prime +e_1\ a_1\ \mathcal{F}(\chi_1,\chi_2)+e_2\ a_2\ \mathcal{G}(\chi_1,\chi_2)+e_3\ a_3\ \mathcal{H}(\chi_3,\chi_4)+e_4\ a_4\ \mathcal{M}(\chi_3,\chi_4).\nonumber\\
\end{eqnarray}
\\
Here, $\chi_1$, $\chi_2$, $\chi_3$ and $\chi_4$ are four angles introduced here solely to help us perform
the simplification method. These angles will be fixed with some angle-fixing ansatzs later. Moreover, $a_i$ are some parameters which, as will be shown later, are related to angular momentum parameters. The following constraints for the functions $\mathcal{F},\mathcal{G}, \mathcal{H}$ and $\mathcal{M}$ are considered:
\begin{eqnarray}
&&\mathcal{F}(\chi_1,\chi_2)=\mathcal{F}(\chi_1),\ \ \ \ \ \mathcal{G}(\chi_1,\chi_2)=\mathcal{G}(\chi_2),\nonumber\\
&&\mathcal{H}(\chi_3,\chi_4)=\mathcal{H}(\chi_3),\ \ \ \mathcal{M}(\chi_3,\chi_4)=\mathcal{M}(\chi_4).
\end{eqnarray}
Based on the transformations (\ref{rutras}) and the above conjectures,

\begin{eqnarray}
&&dr=dr^\prime +e_1\ a_1\ \sin\chi_1\ \sin\chi_2\ d\chi_1-e_2\ a_2\ \sin\chi_1\ \cos\chi_2\ d\chi_2-e_3\ a_3\ \cos\chi_3\ \sin\chi_4\ d\chi_3\nonumber\\
&&-e_4\ a_4\ \cos\chi_3\ \cos\chi_4\ d\chi_4,\nonumber\\
&&du=du^\prime -e_1\ a_1\ \sin\chi_1\ \sin\chi_2\ d\chi_1+e_2\ a_2\ \sin\chi_1\ \cos\chi_2\ d\chi_2+e_3\ a_3\ \cos\chi_3\ \sin\chi_4\ d\chi_3\nonumber\\
&&+e_4\ a_4\ \cos\chi_3\ \cos\chi_4\ d\chi_4.\nonumber\\
\end{eqnarray}
As is clear from Eq. (\ref{rutras}), the transformation takes place simultaneously
in $(r,\phi_i)$ planes. Here, the following angle-fixing ansatzs are introduced:
\begin{eqnarray}\label{ansatz9}
&&\ \ e_1\ d\chi_1=\sin\chi_1\ \sin\chi_2\ d\phi_1,\ \ \ \chi_1\equiv\theta_1, \chi_2\equiv\theta_2=\text{fixed},\nonumber\\
&&-e_2\ d\chi_2=\sin\chi_1\ \cos\chi_2\ d\phi_2,\ \ \ \chi_2\equiv\theta_2, \chi_1\equiv\theta_1=\text{fixed},\nonumber\\
&&-e_3\ d\chi_3=\cos\chi_3\ \sin\chi_4\ d\phi_3,\ \ \ \chi_3\equiv\theta_1, \chi_4\equiv\theta_3=\text{fixed},\nonumber\\
&&-e_4\ d\chi_1=\cos\chi_3\ \sin\chi_4\ d\phi_1,\ \ \ \chi_4\equiv\theta_3, \chi_3\equiv\theta_1=\text{fixed}.\nonumber\\
\end{eqnarray}
\\
The above angle-fixing ansatzs may seem a bit restrictive. One needs to explore the mathematical concept behind such ansatzs. Suppose one fixes $\chi_1\equiv\theta_1$ while $\theta_2$ is restricted to be a fixed angle. Such an assumption means that in the $(r, \phi_1)$ plane one has to perform the rotation with $\theta_2$ to be considered a fixed angle while there are no other restrictions on $\theta_1$ and $\theta_3$. In the $(r, \phi_2)$ plane the rotation has to be performed with a fixed polar angle $\theta_1$. One can continue the discussion for the other two planes. It should be noted that the rotations of the $(r, \phi_i)$ planes are independent of each other. As will be shown later, such angle-fixing ansatzs will generate four independent angular momenta $a_i$, each corresponding to $(r, \phi_i)$ planes.
\\
With the new set of transformations introduced in (\ref{rutras}), the metric (\ref{null9dm}) takes the form
\begin{eqnarray}\label{new9}
&&ds^2=-(du^\prime -e_1\ a_1\ \sin\chi_1\ \sin\chi_2\ d\chi_1+e_2\ a_2\ \sin\chi_1\ \cos\chi_2\ d\chi_2\nonumber\\
&&+e_3\ a_3\ \cos\chi_3\ \sin\chi_4\ d\chi_3+e_4\ a_4\ \cos\chi_3\ \cos\chi_4\ d\chi_4)\nonumber\\
&& [(du^\prime -e_1\ a_1\ \sin\chi_1\ \sin\chi_2\ d\chi_1+e_2\ a_2\ \sin\chi_1\ \cos\chi_2\ d\chi_2\nonumber\\
&&+e_3\ a_3\ \cos\chi_3\ \sin\chi_4\ d\chi_3+e_4\ a_4\ \cos\chi_3\ \cos\chi_4\ d\chi_4)\nonumber\\
&&+2(dr^\prime +e_1\ a_1\ \sin\chi_1\ \sin\chi_2\ d\chi_1-e_2\ a_2\ \sin\chi_1\ \cos\chi_2\ d\chi_2\nonumber\\
&&-e_3\ a_3\ \cos\chi_3\ \sin\chi_4\ d\chi_3-e_4\ a_4\ \cos\chi_3\ \cos\chi_4\ d\chi_4)]\nonumber\\
&&+(1-\tilde{f}(r)) (du^\prime -e_1 a_1\sin\chi_1\ \sin\chi_2\ d\chi_1+e_2 a_2\sin\chi_1\ \cos\chi_2\ d\chi_2\nonumber\\
&&+e_3\ a_3\cos\chi_3\ \sin\chi_4\ d\chi_3+e_4 a_4\cos\chi_3\ \cos\chi_4\ d\chi_4)^2 \nonumber\\
&&+\text{angular parts}.
\end{eqnarray}
It should be noted that $\tilde{f}(r)$ is the transformed form of the function $f(r)$. It will be discussed later in this section. The angular part of the above metric consists of five main parts as follows.
\\
\\
1. The first part is $r^2 d\mu_i^2=d\theta_1^2+\sin^2\theta_1\ d\theta_2^2+\cos^2\theta_1\ d\theta_3^2$. The following transformation rule is adopted:
\begin{eqnarray}\label{newtransf}
&&r^2\ \sum_{i=1}^4\ d\mu_i^2\rightarrow r^2\ \sum_{i=1}^4\  d\mu_i^2+\sum_{l=1}^4\ \ a_l^2\  \sum_{i=1}^{3} \sum_{j=1}^{3} \  \frac{d\mu_l}{d\theta_i}\ \frac{d\mu_l}{d\theta_j}\ \nonumber\\
&&\equiv \sum_{i=1}^{4} \ (r^2+ a_i^2)\ d\mu_i^2.
\end{eqnarray}

2. The second term is $r^2\sin^2\theta_1\sin^2\theta_2\ d\phi_1^2$.

\begin{eqnarray}
&&r\ d\phi_1=r \ \frac{e_1\ d\chi_1}{\sin\chi_1\ \sin\chi_2}=\Bigl(\frac{e_1\cdot r+r\cdot e_1}{2}\Bigl)\ \frac{ d\chi_1}{\sin\chi_1\ \sin\chi_2}\nonumber\\
&&=\frac{e_1\ d\chi_1}{\sin\chi_1\ \sin\chi_2}\ (r^\prime-e_1\ a_1\ \cos\chi_1\ \sin\chi_2).
\end{eqnarray}
Therefore
\begin{eqnarray}\label{phi1}
&&r^2\sin^2\theta_1\sin^2\theta_2\ d\phi_1^2\longrightarrow \sin^2\theta_1\sin^2\theta_2 \ (r\ d\phi_1)(r\ d\phi_1)^\ast\nonumber\\
&&=d\phi_1^2\ \sin^2\theta_1\sin^2\theta_2\ (r^{\prime2}+a_1^2\cos^2\theta_1\ \sin^2\theta_2).
\end{eqnarray}
But the part which results from the metric (\ref{new9}) also gives us the term $a_1^2\ \sin^4\theta_1\sin^4\theta_2\ d\phi_1^2$. Thus, the coefficient of $d\phi_1^2$ term reads as 

\begin{eqnarray}\label{phi1}
r^2\sin^2\theta_1\sin^2\theta_2\ d\phi_1^2\longrightarrow \sin^2\theta_1\sin^2\theta_2 \ (r^{\prime2}+a_1^2\ \sin^2\theta_2)\ d\phi_1^2.
\end{eqnarray}
Indeed we already defined $\theta_2$ to be a fixed angle in the ansatz (\ref{ansatz9}), when we have the rotation in the $(r,\phi_1)$ plane. Hence, the term $\sin^2\theta_2$ is just a fixed real value and can be absorbed in the angular momentum $a_1$, so that
\begin{eqnarray}\label{phi12}
r^2\sin^2\theta_1\sin^2\theta_2\ d\phi_1^2\longrightarrow \sin^2\theta_1\sin^2\theta_2 \ (r^{\prime2}+a^{\prime2})\ d\phi_1^2.
\end{eqnarray}
\\
\\
3. The third term is $r^2 \sin^2\theta_1\ \cos^2\theta_2 \ d\phi_2^2$. Similarly to the transformation of the first term, one has

\begin{eqnarray}\label{phi2}
r\ d\phi_2=\frac{e_2\ d\chi_2}{\sin\chi_1\ \cos\chi_2}\ (r^\prime-e_2\ a_2\ \sin\chi_1\ \sin\chi_2).
\end{eqnarray}
Thus
\begin{eqnarray}\label{phi1}
&&r^2\sin^2\theta_1\cos^2\theta_2\ d\phi_2^2\longrightarrow \sin^2\theta_1\cos^2\theta_2 \ (r\ d\phi_2)(r\ d\phi_2)^\ast\nonumber\\
&&=r^2 \sin^2\theta_1\ \cos^2\theta_2 \ (r^{\prime2}+a_2^2\ \sin^2\theta_1\ \sin^2\theta_2)\ d\phi_2^2 .
\end{eqnarray}
\\
The part which results from the metric (\ref{new9}) also gives the term $a_2^2\ \sin^4\theta_1\cos^4\theta_2 \ d\phi_2^2$. Thus, the coefficient of $d\phi_2^2$ reads as 

\begin{eqnarray}\label{phi2}
r^2\sin^2\theta_1\cos^2\theta_2\ d\phi_2^2\longrightarrow \sin^2\theta_1\cos^2\theta_2 \ (r^{\prime2}+a_2^2\ \sin^2\theta_1)\ d\phi_1^2.
\end{eqnarray}
As we already defined $\theta_1$ to be a fixed angle in the ansatz (\ref{ansatz9}), when we have the rotation in the $(r,\phi_2)$ plane, the term $\sin^2\theta_1$ is a fixed real value and can be absorbed in the angular momentum $a_2$, i.e.
\begin{eqnarray}\label{phi22}
r^2\sin^2\theta_1\cos^2\theta_2\ d\phi_2^2\longrightarrow \sin^2\theta_1\cos^2\theta_2 \ (r^{\prime2}+a_2^{\prime2})\ d\phi_2^2.
\end{eqnarray}
\\
\\
4. The fourth term is $r^2\ \cos^2\theta_1\ \sin^2\theta_3\ d\phi_3^2$ which is transformed as
\begin{eqnarray}
r^2\ \cos^2\theta_1\ \sin^2\theta_3\ d\phi_3^2\longrightarrow \cos^2\theta_1\ \sin^2\theta_3\ (rd\phi_3)(rd\phi_3)^\ast.
\end{eqnarray}
Moreover, there is a term which  results from the metric (\ref{new9}) as $r^2\ \cos^4\theta_1\ \sin^4 \theta_3\ d\phi_3^2$. Using the ansatz (\ref{ansatz9}) and the fact that in the rotation of the plane $(r,\phi_3)$, the angle $\theta_1$  is fixed, one has
\begin{eqnarray}\label{phi32}
r^2\ \cos^2\theta_1\ \sin^2\theta_3\ d\phi_3^2\longrightarrow \cos^2\theta_1\ \sin^2\theta_3\ (r^{\prime2}+a_3^{\prime2})\ d\phi_3^2.
\end{eqnarray}
\\
\\
5. Last, the fifth term $r^2\cos^2\theta_1\ \cos^2\theta_3\ d\phi_4^2$ transforms as

\begin{eqnarray}
r^2\cos^2\theta_1\ \cos^2\theta_3 d\phi_4^2\longrightarrow \cos^2\theta_1\ \cos^2\theta_3 (r^{\prime2}+a_4^{\prime2})\ d\phi_4^2,
\end{eqnarray}
where the term resulting from the metric (\ref{new9}) as $r^2\ \cos^4\theta_1\ \cos^4 \theta_3\ d\phi_4^2$ is also added. It has also been considered that in the rotation of the plane $(r,\phi_4)$, the fixed angle is $\theta_3$.

\subsection{Complex transformation of $f(r)$}
As was mentioned before, the associativity property is missed on $S^7$ and in octonion algebra. Therefore, defining the transformation of different powers of $r$ is not unambiguous i.e. $r\cdot r^2 \neq r^2\cdot r$ on $S^7$. We propose the following transformation for $r^3$:

\begin{eqnarray}\label{r3}
&&r^3\rightarrow r^3-e_1^\prime\ r^2 \biggr[e_1\ a_1 \sqrt{1-\mu_1^2}+e_2\ a_2 \sqrt{1-\mu_2^2}+e_3\ a_3 \sqrt{1-\mu_3^2}+e_4\ a_4 \sqrt{1-\mu_4^2}\biggr]\nonumber\\
&&-e_2^\prime\ r\ \biggr[e_1\ a_1\ a_2\ \sqrt{1-\mu_1^2-\mu_2^2}+e_2\ a_1\ a_3\ \sqrt{1-\mu_1^2-\mu_3^2}+e_3\ a_1\ a_4\ \sqrt{1-\mu_1^2-\mu_4^2}\nonumber\\
&&+e_4\ a_2\ a_3\ \sqrt{1-\mu_2^2-\mu_3^2}+e_5\ a_2\ a_4\ \sqrt{1-\mu_2^2-\mu_4^2}+e_6\ a_3\ a_4\ \sqrt{1-\mu_3^2-\mu_4^2}\biggr]\nonumber\\
&&-e_3^\prime\ \biggr[e_1\ a_1\ a_2\ a_3\ \sqrt{(1-\mu_1^2-\mu_2^2-\mu_3^2)}+e_2\ a_1\ a_2\ a_4\ \sqrt{(1-\mu_1^2-\mu_2^2-\mu_4^2)}\nonumber\\
&&+e_3\ a_2\ a_3\ a_4\ \sqrt{(1-\mu_2^2-\mu_3^2-\mu_4^2)}+e_4\ a_1\ a_3\ a_4\ \sqrt{(1-\mu_1^2-\mu_3^2-\mu_4^2)}\biggr].
\end{eqnarray}
\\
This is our prescription for $r^{3}$, and we regard it as acceptable because it leads
to the correct form of the nine-dimensional Myers-Perry solution. Thus,
\begin{eqnarray}
&&\rho^6\equiv r^6+r^4 \Bigr[a_1^2\ (1-\mu_1^2)+a_2^2\ (1-\mu_2^2)+a_3^2\ (1-\mu_3^2)+a_4^2\ (1-\mu_4^2)\Bigr]\nonumber\\
&&+r^2 \Bigr[a_1^2\ a_2^2\ (1-\mu_1^2-\mu_2^2)+a_1^2\ a_3^2\ (1-\mu_1^2-\mu_3^2)+a_1^2\ a_4^2\ (1-\mu_1^2-\mu_4^2)\nonumber\\
&&+a_2^2\ a_3^2\ (1-\mu_2^2-\mu_3^2)+a_2^2\ a_4^2\ (1-\mu_2^2-\mu_4^2)+a_3^2\ a_4^2\ (1-\mu_3^2-\mu_4^2)\Bigr]\nonumber\\
&&+a_1^2\ a_2^2\ a_3^2\ (1-\mu_1^2-\mu_2^2-\mu_3^2)+a_1^2\ a_2^2\ a_4^2\ (1-\mu_1^2-\mu_2^2-\mu_4^2)\nonumber\\
&&+a_2^2\ a_3^2\ a_4^2\ (1-\mu_2^2-\mu_3^2-\mu_4^2)+a_1^2\ a_3^2\ a_4^2\ (1-\mu_1^2-\mu_3^2-\mu_4^2)\nonumber\\
&&+a_1^2\ a_2^2\ a_3^2\ a_4^2\ (1-\mu_1^2-\mu_2^2-\mu_3^2-\mu_4^2),
\end{eqnarray}
\\
where the last term vanishes because $\sum_{i=1}^4\mu_i^2=1$.\\
Therefore, the transformed $f(r)$, denoted by $\tilde{f}(r)$, reads as 
\begin{eqnarray}
\tilde{f}(r)=1-\frac{2m}{(\rho(r))^6}.
\end{eqnarray}

\subsection{Reverting to the Boyer-Lindquist coordinates}
We can now obtain the transformed metric under the octonions
complexification process in the form (all primes are omitted)

\begin{eqnarray}\label{9dEF}
&&ds^2=-du^2-2du\ dr+ (1-\tilde{f}(r)(du-a_1\ \mu_1^2\ d\phi_1-a_2\ \mu_2^2\ d\phi_2-a_3\ \mu_3^2\ d\phi_3-a_4\ \mu_4^2\ d\phi_4)^2\nonumber\\
&&+2a_1\ \mu_1^2\ dr\ d\phi_1+2a_2\ \mu_2^2\ dr\ d\phi_2+2a_3\ \mu_3^2\ dr\ d\phi_3+2a_4\ \mu_4^2\ dr\ d\phi_4\nonumber\\
&&+(r^2+a_1^2)d\mu_1^2+(r^2+a_2^2)d\mu_2^2+(r^2+a_3^2)d\mu_3^2
+(r^2+a_4^2)d\mu_4^2\nonumber\\
&&+(r^2+a_1^2)\ \mu_1^2\ d\phi_1^2+(r^2+a_2^2)\ \mu_2^2\ d\phi_2^2+(r^2+a_3^2)\ \mu_3^2\ d\phi_3^2+(r^2+a_4^2)\ \mu_4^2\ d\phi_4^2.
\end{eqnarray}
\\ 
The metric (\ref{9dEF}) reveals the correct form of the nine-dimensional Myers-Perry metric in Edington-Finklestein coordinates.
The metric can be re-expressed in the Boyer-Lindquist coordinates through the following transformations:
\begin{eqnarray}
&&du=dt-g(r)\ dr,\nonumber\\
&&d\phi_i=d\phi^\prime_i-h_{\phi_i}\ dr,
\end{eqnarray}
where
\begin{eqnarray}
&&g(r)=\frac{\Pi}{\Delta},\nonumber\\
&&h_{\phi_i}=\frac{\Pi}{\Delta}\ \frac{a_i^2}{r^2+a_i^2},\nonumber\\
&&\Delta=\Pi-m\ r^2,\nonumber\\
&&\Pi=\prod_{i=1}^4 (r^2+a_i^2).
\end{eqnarray}
Thus, the Myers-Perry spinning solution with four independent angular momenta in nine dimensions is derived from the static solution by exploiting the Cayley-Dickson algebra in the case $n=3$.

\section{Concluding Remarks}\label{generald}
In this section, we summarise the results derived in the body of the paper. To sum up, by using Cayley-Dickson sequence of algebras simply defined by complex numbers, quaternions, octonions and sedenions together with Janis-Newman algorithm and Giampieri's simplification we have derived the spinning solutions by introducing only one set of transformations from the static ones in odd dimensions. Thus, one can find the relation between the dimension of space-time $d$, the number of independent angular momenta in each odd dimension and the dimension of the Cayley-Dickson algebra. The Cayley-Dickson algebra produces  $D=2^n$ dimensional algebras only for $n=1,2,3,4$. Therefore,
 
 \begin{eqnarray}\label{NdD}
 N=2^{n-1}=2^{\text{log}_2(d-1)-1},
 \end{eqnarray}
only for $d=3,5,9$ and $d=17$. We recall that the maximum number of 
angular momenta in each dimension $N_{\text{max}}$ is the integer 
part of $\frac{d-1}{2}$. It should be emphasised that Eq. (\ref{NdD}) 
can only be used from right to left. The method stops itself and cannot 
be applied if $d=33$. There are two main reasons: First, after $n=4$, 
there is no algebra produced by the Cayley-Dickson approach; second, the 
Hopf fibration of the sphere is only possible for $S^1, S^3,S^7$ and 
finally $S^{15}$ which is obviously the angular part of a 17-dimensional 
solution. Thus, the spinning Myers-Perry solution in 17 dimensions can be 
derived by using the sedenion algebra (Appendix \ref{sedenions}). Therefore, 
there is a natural cut-off for using Janis-Newman algorithm and the 
Cayley-Dickson algebras if one wants to derive a spinning solution in one 
fell swoop.  Although even dimensions such as $d=4$ were not discussed in 
this paper, the Kerr metric, for instance, can be derived from the static 
solution by using Janis-Newman algorithm and the Cayley-Dickson algebra with 
$n=1$. Our main purpose was to emphasise the connection of the Hopf fibration 
and the Cayley-Dickson construction while using the Janis-Newman approach 
for generating spinning solutions. Moreover, derivation of the electromagnetic 
self-force acting on a static charged particle is possible by applying the 
Janis-Newman algorithm on the self-force in static space-time, an idea presented in 
\cite{baracoli,Nadi}. This approach cannot be used in higher dimensions when 
the Cayley-Dickson construction  with $n\geq3$ occurs, because the transformations 
of different powers of the coordinate $r$ are not unambiguous. The underlying reason 
is that the algebra after $n\geq3$, looses the property of associativity.

\section*{Acknowledgements} 
We would like to thank Machiko Hatsuda for useful
correspondence on the Hopf fibration.
\appendix
\section{Sedenions}\label{sedenions}
Sedenions algebra is a Cayley-Dickson algebra with $n=4$ dimensions that is known as a  non-associative, non-commutative, and non-alternative algebra. The basis elements of sedenions are denoted by $\{e_0, e_1,..., e_{15}\}$, where $e_0$ is the unit elements. A sedenion $s$ can be written as 
\begin{eqnarray}
s=\sum_{i=0}^{15} a_i\ e_i,
\end{eqnarray}
where $a_0,a_1,...,a_{15}$ are real numbers.The sedenions algebra $\mathbb{S}$ is then defined as \cite{Bilgici,Imeada}
\begin{eqnarray}\label{sedenionalgebra}
s=(o_1; o_2) \in \mathbb{S}, \ \ \ o_1, o_2 \in \mathbb{O},
\end{eqnarray}
where $\mathbb{O}$ is the octonion algebra. Thus, a sedenion is an ordered pair of two octonions. The conjugate of a sedenion $s$ is defined as $s=(o_1; -o_2)$.  The product of two sedenions is 
\begin{eqnarray}
s_1 s_2=(o_1 o_2+\rho \bar{o}_4 o_2; o_2 \bar{o}_3+o_4 o_1),
\end{eqnarray}
with $\rho$ chosen to be $\rho=-1$.

\end{document}